\documentclass[twocolumn,floats,floatfix,nofootinbib,10pt] {revtex4-1}
\usepackage{graphicx}
\usepackage{dcolumn}
\usepackage{bm}
\usepackage{graphics}
\usepackage{slashed}
\usepackage{amssymb}
\usepackage{natbib}
\usepackage{amsmath}
\usepackage{url}
\usepackage[usenames,dvipsnames,svgnames,table]{xcolor}
\usepackage{lipsum}
\usepackage{color}

\begin{document}

\title{From Hopf fibrations to exotic causal replacements}

\author{Miguel Bezares}
\affiliation{ Facultad de Ciencias de la Educaci\'on,
Escuela de Matem\'aticas, Universidad San Sebastian, Bellavista 7, Santiago 8420524, Chile}
\affiliation{ Departament de F\'isica Universitat de les Illes Balears and Institut d'Estudis Espacials de Catalunya, Palma de Mallorca, Baleares E-07122, Spain}
\author{\'Erico Goulart}
\email{Corresponding author. E-mail: egoulart@cbpf.br}
\affiliation{Centro Brasileiro de Pesquisas F\'isicas - CBPF, Rua Dr. Xavier Sigaud, 150, Urca, CEP 22290-180, Rio de Janeiro, Brazil}
\author{Gonzalo Palomera}
\affiliation{Facultad de Ingenier\'ia, Universidad del Desarrollo, Santiago 7620001, Chile}
\author{Daniel J. Pons}
\email{E-mail: dpons@unab.cl}
\affiliation{Facultad de Ciencias Exactas, Departamento de Matem\'aticas, Universidad Andres Bello, Rep\'ublica 220, Santiago, Chile}
\author{Enrique G. Reyes}
\email{E-mail: e\_g\_reyes@yahoo.ca ; enrique.reyes@usach.cl}
\affiliation{Departamento de Matem\'atica y Ciencia de la Computaci\'on Universidad de Santiago de Chile Casilla 307 Correo 2, Santiago, Chile}
 
\begin{abstract}
Topological solitons are relevant in several areas of physics \cite{Manton}. Recently, these configurations have been investigated in contexts as diverse as hydrodynamics \cite{Klech}, Bose-Einstein condensates \cite{Hall}, ferromagnetism \cite{Coo}, knotted light \cite{Irv} and non-abelian gauge theories \cite{Fad}. In this paper we address the issue of wave propagation about a static Hopf soliton in the context of the Nicole model. Working within the geometrical optics limit we show that several nontrivial lensing effects emerge due to nonlinear interactions as long as the theory remains hyperbolic. We conclude that similar effects are very likely to occur in effective field theories characterized by a topological invariant such as the Skyrme model of pions.
\end{abstract}

\maketitle

\section{Introduction}

\quad\ Start with a rectangular rubber sheet and draw a collection of non-intersecting horizontal lines on it. Make sure you use a thin tip so as to fill the material with as many lines as possible. Glue the top edge of the rectangle to its bottom and give a full twist to the cylinder before you paste its left and right hand sides to form a torus. Now take the toroid structure and look at the original lines. Somehow they became linked! Repeat this recipe with slightly bigger (and also smaller) rectangles and try to assemble the resulting torii one inside the other, very much in the same way as  matryoshka nested dolls are constructed. If you are patient enough to perform this continuously, you'll end up with a fibered structure filling most of $\mathbb{R}^{3}$ which captures the essence of the Hopf fibration \cite{Seven}.

In mathematics, the Hopf fibration \cite{Hopf} describes the 3-sphere $\mathbb{S}^{3}$ in terms of a disjoint union of circles $\mathbb{S}^{1}$ and an ordinary 2-sphere $\mathbb{S}^{2}$, with fiber structure
\begin{equation}
 \mathbb{S}^{1}\hookrightarrow\mathbb{S}^{3}\xrightarrow{\pi}\mathbb{S}^{2},
\end{equation}
$\pi$ denoting the projection map. In one of its simpler forms, $\pi$ may be defined as follows. Identify $\mathbb{S}^{3}$ with the subset $(z_{0},z_{1})\in\mathbb{C}^{2}$, such that $|z_{0}|^{2}+|z_{1}|^{2}=1$, and $\mathbb{S}^{2}$ with the subset $(z,x)\in\mathbb{C}\times\mathbb{R}$, such that $|z|^{2}+x^{2}=1$. Then the projection map reads 
\begin{equation}
\pi(z_{0},z_{1})=(2z_{0}z_{1}^{*},|z_{0}|^{2}-|z_{1}|^{2}).
\end{equation}
It turns out that a point $(w_{0},w_{1})=(\lambda z_{0},\lambda z_{1})$, with $\lambda\in\mathbb{C}$, will be mapped to the same point of $\mathbb{S}^{2}$ if and only if $|\lambda|^{2}=1$, thus generating the structure of circular fibers embedded in $\mathbb{S}^{3}$. The remarkable structure of $\mathbb{R}^{3}$ filled with nested torii made of linked circles described above appears when we compose $\pi$ with stereographic projection of $\mathbb{S}^{3}$ onto $\mathbb{R}^{3}$. It can be shown \cite{projection} that such stereographic projection maps the Hopf fibers to linked circles in $\mathbb{R}^{3}$, the only exception being the Hopf circle containing the projection point which is mapped to a straight line in $\mathbb{R}^{3}$, a ``circle through infinity". 

This paper deals with wave propagation in a nonlinear field theory --the so called Nicole model \cite{Nicole,Radu}-- determining nontrivial fibrations of $\mathbb{R}^{3}$ of the Hopf type induced by exact finite-energy solutions to the equations of motion. For the Nicole model, the projection map emerges as the critical point of a Lorentz invariant Lagrangian theory of maps from flat space-time into the 2-sphere, or equivalently into $SU(2)/U(1)$\footnote{In order to distinguish between the former Hopf map $\pi$ and its spacetime version considered here, we shall denote the latter by $\varphi:\mathbb{R}^{1+3}\rightarrow\mathbb{S}^{2}$ and, following references \cite{Adam,Gil}, call it a Hopf map as well.}. The Lagrangian is a simple non-polynomial generalization of the $O(3)$ $\sigma$-model \cite{Tataru}, and the associated Euler-Lagrange equations are given by a system of second order quasi-linear PDE's for the map, see equation \eqref{EOM} below. Nicole's motivation was to improve earlier attempts \cite{Enz} at constructing extended solutions of the `twisted ring' type in three spatial dimensions.  In order to achieve this, Nicole modified the action appearing in \cite{Tataru} in an exotic fashion so as to give a scaling neutral theory \cite{Derrick}, very much in the same way as was done before by Deser \textit{et al} in \cite{Deser}. Nowadays, similar solutions are often discussed in the literature in various contexts ranging from the models of condensed matter physics \cite{Baba,Met} to high energy physics and cosmology \cite{Bran, Davis, Vil, AFZ}. 
 
The main question we address here is the following: How do small perturbations about a given Hopf map propagate in space-time, according to Nicole's equations? Working within the geometrical optics limit we shall show that several lensing effects emerge, yielding an exotic causal replacement governing the kinematics of perturbations. Using the effective metric technique (see, for instance, \cite{Viss}), we work out numerically several interesting situations and discuss how the topological charge may affect the behaviour of rays. In this sense, the present work is a natural generalization of previous results sketched in \cite{Er1}.

\section{The model}

\subsection{Kinematics}

\quad\ We write $(\textbf{M},\ g)$ for the $1+3$ dimensional Minkowski spacetime with metric signature $(-,+,+,+)$ and $(\mathbb{S}^{2},\ h)$ for the unit 2-sphere with metric $h$. Inspired by the classical paper  \cite{Eells}, we are interested in surjective maps
\begin{equation}\label{mapping}
\varphi: (\textbf{M},\ g) \rightarrow (\mathbb{S}^{2},\ h)\; ,
\end{equation}
see also the interesting review  \cite{Bruhat}.
If $x^{a}$ $(a=0,1,2,3)$ and $y^{\alpha}$  $(\alpha=1,2)$ denote local coordinates in the base and target spaces, respectively, the map $\varphi$ reads
\begin{equation}
y^{\alpha}=\varphi^{\alpha}(x^{a}).
\end{equation}
The differential of $\varphi$ at $x\in \textbf{M}$ is the best linear approximation of $\varphi$ near $x$
\begin{equation}  
d\varphi_{x}: T_{x}\textbf{M}\rightarrow T_{\varphi(x)}\mathbb{S}^{2},
\end{equation}
and we use it to pullback tensors living on $\mathbb{S}^{2}$ to the spacetime $\textbf{M}.$ In particular, we write
\begin{eqnarray}
(\varphi^{*}h)_{ab}\equiv L_{ab}= h_{\alpha\beta}\partial_{a}\varphi^{\alpha}\partial_{b}\varphi^{\beta}
\end{eqnarray}
and 
\begin{eqnarray}
(\varphi^{*}\epsilon)_{ab}\equiv F_{ab}=\epsilon_{\alpha\beta}\partial_{a}\varphi^{\alpha}\partial_{b}\varphi^{\beta},
\end{eqnarray}
with $h_{\alpha\beta}$ and $\epsilon_{\alpha\beta}$ denoting the metric and area 2-form to the unit sphere, respectively. We shall call $\varphi^{*}h$ the \textit{strain for the map} $\varphi$ and $\varphi^{*}\epsilon$ the \textit{Hopf curvature}.

In physical applications, it is natural to consider configurations such that the field tends to a single value as spatial infinity is approached in any direction. We choose this `constant state' such that `infinity' corresponds to the north pole $\mathcal{N}$ on the 2-sphere. This procedure is effectively equivalent to a one-point compactification of $\mathbb{R}^{3}$, which we denote by $\mathbb{R}^{3}_{0}$. Since the compactified three-dimensional Euclidean space $\mathbb{R}^{3}_{0}$ is topologically equivalent to a topological three-sphere $\mathbb{S}^{3}$ we have, at any given time a function 
\begin{equation}
\varphi_{t}:\mathbb{R}^{3}_{0}\cong\mathbb{S}^{3}\rightarrow \mathbb{S}^{2}\; .
\end{equation}
Now, as the third homotopy group of the target space is $\pi_{3}(\mathbb{S}^{2})\in\mathbb{Z}$, there is an associated integer topological charge $Q$ called the \textit{Hopf charge}. As a consequence, the possible mappings (\ref{mapping}) fall into different homotopy classes, and a \textit{Hopf soliton} is a configuration that is a critical point for an energy functional (see subsection II.B) within a fixed homotopy class. 

Whitehead first showed in reference \cite{White} that it is possible to express the Hopf charge $Q$ as an integral of the form 
\begin{equation}\label{topo}
 Q=\frac{1}{16\pi^{2}} \int_{\mathbb{R}^3}  \stackrel{\ast}{F_{ab}}C^{a}t^{b} d^{3}x\; ,
\end{equation}
in which $\stackrel{\ast}{F_{ab}}\equiv\frac{1}{2}\eta^{abcd}F_{cd}$ is the dual tensor to the Hopf curvature, $t^{b}$ is a normalized timelike vector field ($t^{a}t_{a}=-1$) orthogonal to the space slices and (using that $F_{ab}$ is a closed 2-form) we define $C^{a}$ locally via $F_{ab}=\partial_{[a}C_{b]}$. Roughly speaking, $Q$ will remain invariant under arbitrary smooth deformations of the map and, in particular, under time evolution. Interestingly, for a given time $t$, the preimage $\varphi^{-1}_{t}$ of a point $P\in\mathbb{S}^{2}$ will be, in general, a closed loop in $\mathbb{R}^{3}_{0}\cong\mathbb{S}^{3}$. Heuristically, we can compute $Q$  by counting the linking number between two such preimage curves. We define the position of the Hopf soliton as the preimage of the south pole $\mathcal{S}\in\mathbb{S}^{2}$, as it corresponds to the position in $\mathbb{S}^{2}$ which is the most distant to the vacuum. 

 \subsection{Dynamics}

Nicole's model \cite{Nicole} consists of maps which are stationary points of the action
\begin{equation}
S[\varphi]=\int_{\textbf{M}}\sigma_{1}^{3/2}\ dv_{g},
\end{equation}
where $dv_{g}$ is the volume element determined by $g$ and $\sigma_{1}=L^{c}_{\phantom a c}$ is the first elementary symmetric polynomial constructed with the pulled-back metric $\varphi^{*}h$. For the sake of conciseness we write henceforth $\mathcal{L}$ for the Lagrangian density, $\mathcal{L}_{1}=\partial{\mathcal{L}}/\partial\sigma_{1}$ and $\mathcal{L}_{11}=\partial^{2}{\mathcal{L}}/\partial\sigma_{1}^{2}$. With these conventions, the Euler-Lagrange equations become
\begin{equation}\label{trace}
\mbox{trace}_{g}\big[\mathfrak{D}\big(\mathcal{L}_{1}d\varphi\big)\big]=0,
\end{equation}
with $\mathfrak{D}$ the linear connection in the associated vector bundle $\textbf{E}=T^{*}\textbf{M}\otimes \varphi^{-1}T\mathbb{S}^{2}$ over $\textbf{M}$ (see \cite{Misner}). Written in a local chart, equation (\ref{trace}) becomes 
\begin{eqnarray}\label{EOM}
&&\frac{1}{\sqrt{-g}}\partial_{a}\big(\sqrt{-g}\ \mathcal{L}_{1}\ g^{ab}\partial_{b}\varphi^{\alpha}\big)+\\\nonumber
&&\quad\quad\quad+\Gamma^{\alpha}_{\phantom a \beta\gamma}\ \mathcal{L}_{1}\ \partial^{a}\varphi^{\beta}\partial_{a}\varphi^{\gamma}=0,
\end{eqnarray}
where  sum over repeated indices is understood, and $\Gamma^{\alpha}_{\phantom a \beta\gamma}$ denotes the Christoffel symbols of the Levi-Civita connection on $\mathbb{S}^{2}$.  

It is clear from \eqref{EOM} that the dynamics is given by a 2-dimensional system of quasilinear PDE's for the map (\ref{mapping}). Eqs. (\ref{EOM}) become, after some simple manipulations,
\begin{equation}\label{JA}
M^{ab}_{\phantom a\phantom a  \alpha\beta}(\varphi,\partial\varphi)\ \partial_{a}\partial_{b}\varphi^{\beta}+J_{\alpha}(\varphi,\partial\varphi)=0,
\end{equation}
where $J_{\alpha}$ stands for semilinear terms in $\varphi$ and $M^{ab}_{\phantom a\phantom a  \alpha\beta}$ is the \textit{principal part} of the system. As it is well known, the highest-order terms in derivatives almost completely controls the qualitative behaviour of solutions of a partial differential equation \cite{Cour}. We obtain
 \begin{equation}\label{M}
M^{ab}_{\phantom a\phantom a  \alpha\beta}= g^{ab}h_{\alpha\beta}+\xi h_{\alpha\mu}h_{\beta\nu}\partial^{(a}\varphi^{\mu}\partial^{b)}\varphi^{\nu},
\end{equation}
where $\xi\equiv 2\mathcal{L}_{11}/\mathcal{L}_{1}$ and $(a,b)=(ab+ba)/2$.  
 
\section{Regular Hyperbolicity}
 
A key issue to elucidate about systems of the form (\ref{EOM}) is whether they admit a well-posed Cauchy problem. In other words, we wish to determine the map $\varphi(x^{a})$ --at least for some finite interval of time $T$-- if initial data $\varphi|_{\Sigma}$, $\partial\varphi|_{\Sigma}$ are given in a non-characteristic hyper-surface $\Sigma\subseteq\textbf{M}$. In order to guarantee well-posedness for the Nicole model it is convenient to work in the framework of regular hyperbolicity introduced by Christodoulou \cite{Christ} (see also \cite{Wong} for a short review of  Christodoulou's work, and \cite{Kamran} for a geometric approach to classical hyperbolicity). Roughly speaking, hyperbolicity is an algebraic property of the principal part $M^{ab}_{\phantom a\phantom a \alpha\beta}$ given by \eqref{M}, entailing the existence of solutions for arbitrary smooth initial data, and uniqueness and continuous dependence on the initial data.  

If we are given $\varphi$ and $\partial\varphi$ at a spacetime point $x$ we can evaluate the principal part \eqref{M} corresponding to that point. Formally, we have the map between the fibre bundles
\begin{equation}\nonumber
M^{ab}_{\phantom a\phantom a \alpha\beta}(x):T_{x}^{*}\textbf{M}\otimes T_{x}^{*}\textbf{M}\rightarrow T^{*}_{\varphi(x)}\mathbb{S}^{2}\otimes T^{*}_{\varphi(x)}\mathbb{S}^{2}.
\end{equation}
For the sake of conciseness we use the notation
\begin{equation}
M_{\alpha\beta}(x,\eta):= M^{ab}_{\phantom a\phantom a \alpha\beta}(x)\eta_{a}\eta_{b},
\end{equation}
for an arbitrary covector $\eta\in T^{*}_{x}\textbf{M}$. This object is called the \textit{principal symbol} and, according to Christodoulou \cite{Christ}, the system will be regular hyperbolic if:
\begin{itemize}
\item{There exists a scalar function $t(x)$ such that $M_{\alpha\beta}(x,dt)$ is negative definite;}
\item{There exists $X^{a}(x)$ such that $M_{\alpha\beta}(x,\eta)$ is positive definite for all $\eta$ satisfying $X^{a}\eta_{a}$=0;}
\end{itemize}
Functions $t(x)$ are called \textit{time functions} while vectors $X^{a}(x)$ are called \textit{observer fields}. The existence of these quantities is sufficient to construct algorithmically a \textit{compatible energy current} $J^{a}(X,\varphi,\partial\varphi,...,\partial^{k}\varphi)$ in order to apply the energy estimates. The latter imply that solutions exist and depend continuously on the data given in a level set $\Sigma_{t}$ of $\textbf{M}$. Furthermore, one can also guarantee that the evolution of small disturbances (linearized waves) about some given smooth data is properly-posed for the corresponding linearized equations.

In order to check the existence of $t(x)$ and $X^{a}(x)$ for the symbol (\ref{M}) we use the mixed tensor $M:=h^{\alpha\gamma}M_{\gamma\beta}$. Its eigenvalues satisfy the second order algebraic equation
\begin{equation}\label{eigen}
\lambda^{2}-\mbox{Tr} M\lambda+\mbox{det} M=0,
\end{equation}
at each $x\in M.$ Interestingly, for the symbol (\ref{M}), the trace and the determinant are real-valued functions of $x$ and $\eta$ which are always given in terms of two quadratic forms in the cotangent bundle $T^{*}\textbf{M}$
\begin{eqnarray}\label{p1p2}
&&\mbox{Tr} M=\mathcal{P}_{1}(x,\eta)+\mathcal{P}_{2}(x,\eta)\\
&&\mbox{det} M=\mathcal{P}_{1}(x,\eta)\mathcal{P}_{2}(x,\eta),
\end{eqnarray}
with
\begin{eqnarray}\label{reci}
&&\mathcal{P}_{1}(x,\eta)=g^{ab}(x)\eta_{a}\eta_{b}\\\label{reci1}
&&\mathcal{P}_{2}(x,\eta)=(m^{-1})^{cd}(\varphi(x))\eta_{c}\eta_{d}
\end{eqnarray}
and
\begin{equation}\label{reciprocal}
(m^{-1})^{ab}:= g^{ab}+\xi L^{ab}.
\end{equation}
Moreover, a closer inspection of (\ref{eigen}) gives:
\begin{equation}
\lambda_{\pm}(x,\eta)=\frac{\mathcal{P}_{1}+\mathcal{P}_{2}}{2}\pm\frac{|\mathcal{P}_{1}-\mathcal{P}_{2}|}{2}.\label{deter}
\end{equation}

In what follows we shall use the terminology \textit{reciprocal effective metric} for the quantity $m^{-1}$ defined in (\ref{reciprocal}).  

The system will be regular hyperbolic if the  reciprocal effective metric satisfies some algebraic conditions. 
An inescapable (necessary) condition for the system to be well-posed is that $m^{-1}$ constitutes a non-degenerate semi-definite tensor field with a Lorentzian signature. Since $m^{-1}$ depends explicitly on the solution to (\ref{EOM}) we are considering, we naturally expect these conditions to imply that not all initial data are admissible for the model, as we now show.

Following Manton \cite{Nick} we suppose that $L_{ab}$ can be diagonalized relative to $g_{ab}$ in a given point. The eigenvalues of $L_{ab}$ are necessarily nonnegative and due to rank considerations two of them must vanish identically. We write
\begin{equation}
L_{ab}=(\lambda_{0}^{2},\lambda_{1}^{2},\lambda_{2}^{2},\lambda_{3}^{2})
\end{equation}
It follows
\begin{eqnarray*}
&&(m^{-1})^{00}=\frac{2\lambda_{0}^{2}-\lambda_{1}^{2}-\lambda_{2}^{2}-\lambda_{3}^{2}}{-\lambda_{0}^{2}+\lambda_{1}^{2}+\lambda_{2}^{2}+\lambda_{3}^{2}}\\
&&(m^{-1})^{11}=\frac{-\lambda_{0}^{2}+2\lambda_{1}^{2}+\lambda_{2}^{2}+\lambda_{3}^{2}}{-\lambda_{0}^{2}+\lambda_{1}^{2}+\lambda_{2}^{2}+\lambda_{3}^{2}}\\
&&(m^{-1})^{22}=\frac{-\lambda_{0}^{2}+\lambda_{1}^{2}+2\lambda_{2}^{2}+\lambda_{3}^{2}}{-\lambda_{0}^{2}+\lambda_{1}^{2}+\lambda_{2}^{2}+\lambda_{3}^{2}}\\
&&(m^{-1})^{33}=\frac{-\lambda_{0}^{2}+\lambda_{1}^{2}+\lambda_{2}^{2}+2\lambda_{3}^{2}}{-\lambda_{0}^{2}+\lambda_{1}^{2}+\lambda_{2}^{2}+\lambda_{3}^{2}}.
\end{eqnarray*}
We then conclude that the reciprocal effective metric is Lorentzian if its eigenvalues satisfy the inequality
\begin{equation}
\lambda_{0}^{2}<\frac{\lambda_{1}^{2}+\lambda_{2}^{2}+\lambda_{3}^{2}}{2}.
\end{equation}
In particular, that will be true if $\lambda_{0}^{2}=0$, as is the case for all static solitonic solutions. As we are mainly interested in wave propagation about these solutions, we assume henceforth that $m^{-1}$ is Lorentzian. 

\section{Causal Replacements}

By a causal replacement we mean the fact that linearized waves propagating about a smooth background solution $\varphi_{0}$ do not travel with the velocity of light. For the Nicole model, as for quasi-linear hyperbolic field theories in general, wave propagation depends on the particular solution, direction of propagation and `wave polarization' as explained, for instance, in the reference \cite{tan}. Accordingly, in order to analyze the causal replacement of our model, we recall the notion of characteristics. Physically, they can be identified with the infinite-momentum limit of the \textit{eikonal} approximation \cite{Perlick} (or, equivalently, with the surfaces of discontinuity obtained via Hadamard's method). 

A hyper-surface $\Sigma\subseteq\textbf{M}$, given by $f(x^{a})=const$, is called characteristic if
\begin{equation}
\mathcal{P}(x,k):=|M_{\alpha\beta}(\varphi_{0}(x),k)|=0,
\end{equation}
with $k_{a}:=\partial_{a}f$. The set
\begin{equation}\label{Cest}
\mathcal{C}_{x}^{*}:=\{k\in T_{x}^{*}\textbf{M}\ \big{|}\ \mathcal{P}(x,k)=0,k\neq 0\},
\end{equation}
is called the characteristic set and it consists of the locus of normal covectors $k$ to the characteristic surfaces at $x$. Roughly, the existence of characteristics in a region implies that linearized waves have a well-behaved, finite velocity about the background solution.

Using (\ref{p1p2}), we obtain
\begin{equation}\label{Fres}
\mathcal{P}(x,k)=\mbox{det}(h_{AB})\mathcal{P}_{1}(x,k)\mathcal{P}_{2}(x,k),
\end{equation}
with $\mathcal{P}_{1}$ and $\mathcal{P}_{2}$ given by (\ref{reci}) and (\ref{reci1}). As a consequence, the wave normals covectors $k$ satisfying \eqref{Cest} are determined by the vanishing sets of a multivariate polynomial of fourth order in $k_{a}\in T^{*}_{x}\textbf{M}$. Thus, the resulting algebraic variety is always given by a product of quadrics, one of which  changes from point to point in a way completely prescribed by the background solution $\varphi_{0}(x)$ and the nonlinearities present in the Nicole model ($\mathcal{P}_{2}=0$).  

As we assume that $(m^{-1})^{ab}$ is non-degenerate, we can always define its inverse $m_{ab}$ i.e., $(m^{-1})^{ac}m_{cb}=\delta^{a}_{\phantom a b}$. The theory of PDE's then proceeds by telling us that the characteristic surfaces themselves are given by the zeros of a dual polynomial in the tangent space $T_{x}\textbf{M}$ which also factorizes, i.e.
\begin{equation}
\mathcal{C}_{x}:=\{q\in T_{x}\textbf{M}\ \big{|}\ \mathcal{P}^{\#}(x,q)=0,q\neq 0\},
\end{equation}
with
\begin{equation}
\mathcal{P}^{\#}(x,q)=\mathcal{P}^{\#}_{1}(x,q)\mathcal{P}^{\#}_{2}(x,q),
\end{equation}
and $\mathcal{P}^{\#}_{1}(x,q)=g_{ab}(x)q^{a}q^{b}$, $\mathcal{P}^{\#}_{2}(x,q)=m_{ab}(\varphi_{0}(x))q^{a}q^{b}$. Thus, the model supports two types of waves: one is governed by the background metric $g_{ab}$ while the other is governed by the \textit{effective metric} $m_{ab}(\varphi_{0}(x))$. Note, however, that, since the background solutions carry a topological index Q given by (\ref{topo}), the effective metric will also depend implicitly on $Q$.
 
It is well known that when $m_{ab}$ is Lorentzian, the vectors $q^{a}$ such that $\mathcal{P}_{2}^{\#}(x,q)=0$ (in a region of spacetime) satisfy the equation of null geodesics w.r.t. the effective metric, see \cite{Nov}, i.e.
\begin{equation}\label{effgeo}
q^{a}_{\phantom a || b}q^{b}=0
\end{equation}
where $||$ denotes covariant differentiation w.r.t. $m_{ab}$ (see, however, \cite{Er2} for a case where this property fails). Conversely, we are led to investigate spacetime trajectories such that the \textit{effective line element} vanishes, i.e.
\begin{equation}\label{effconst}
d\hat{s}^{2}=m_{ab}(\varphi_{0}(x))dx^{a}dx^{b}=0.
\end{equation}
Interestingly, if the map (\ref{mapping}) is independent of time, the problem of finding null geodesics in the effective space-time reduces to that of finding geodesics in an effective Riemannian manifold of lower dimension. Indeed, choosing a coordinate system $x^{a}=(t,x^{i})$, $i=1,2,3$, such that $g_{0i}=0$ and $\partial_{t}\varphi^{\alpha}=0$ one reduces (\ref{effgeo}) and (\ref{effconst}) to the equations
\begin{equation}\label{ODE}
\ddot{x}^{i}+\hat{\Gamma}^{i}_{\phantom a jk}\dot{x}^{j}\dot{x}^{k}=0
\end{equation}
\begin{equation}\label{CONS}
m_{ij}(\varphi_{0}(x))\dot{x}^{i}\dot{x}^{j}=1,
\end{equation}
with $\dot{x}^{i}:=\frac{dx^{i}}{dt}$ and $\hat{\Gamma}^{i}_{\phantom a jk}$ the Christoffel symbols associated with the spatial part of the effective metric. Therefore, we can qualitatively describe the interaction between background Hopfions and rays by computing the geodesics of an effective three-dimensional manifold (see \cite{Low} for a discussion in the context of the Schwarzschild solution).

 \section{Static Hopf solitons}

We briefly review here how solutions of Eq. (\ref{EOM}) with a nonvanishing Hopf index emerge in the simpler case of static maps (see \cite{Gil} for more details). We start by introducing toroidal coordinates $(\eta,\theta,\psi)$ in $\mathbb{R}^{3}$:
\begin{eqnarray*}
x&=&q^{-1}\ \mbox{sinh}\eta\ \mbox{cos}\psi\\
y&=&q^{-1}\ \mbox{sinh}\eta\ \mbox{sin}\psi\\
z&=&q^{-1}\ \mbox{sin}\theta,
\end{eqnarray*}
with $q:= \mbox{cosh}\eta-\mbox{cos}\theta$, $\eta\in[0,\infty)$, $\theta\in[0,2\pi)$ and $\psi\in[0,2\pi)$. Note that surfaces of constant $\eta$ are given by non-intersecting toroids of different radii:
\begin{equation}
x^{2}+y^{2}+z^{2}+1=2\ \mbox{coth}\eta\ (x^{2}+y^{2})^{1/2}.
\end{equation}
In particular, as $\eta\rightarrow \infty$ the tori asymptotically approach the ring $z=0$, $x^{2}+y^{2}=1$. Conversely, as $\eta\rightarrow 0$ the tori become infinitely large, approaching asymptotically the line through the origin $x^{2}+y^{2}=0$. In these coordinates, the space-time line element has the form
 \begin{equation}\nonumber
ds^{2}=-dt^{2}+q^{-2}\big(d\eta^{2}+d\theta^{2}+\mbox{sinh}^{2}\eta\ d\psi^{2}\big)
\end{equation}

Now we consider the target manifold $(\mathbb{S}^{2}, h)$. Calculations become easier if we use coordinates $(R,\Phi)$ in $\mathbb{S}^{2}$, which can be identified via stereographic projection from the south pole $\mathcal{S}$ to the equatorial plane. In these coordinates, the curves $(R=const_{1})$ describe lines of constant latitude (also called parallels) on the sphere while the curves $(\Phi=const_{2})$ describe its meridians. Particularly, $const_{1}=0$ corresponds to $\mathcal{N}$ while $const_{1}\rightarrow\infty$ corresponds to $\mathcal{S}$. The line element has the simple form
\begin{equation}\nonumber
d\ell^{2}=\frac{4}{(1+R^{2})^{2}}\big(dR^{2}+R^{2}d\Phi^{2}\big),
\end{equation}
and the non-vanishing Chirstoffel symbols are
\begin{eqnarray*}
&&\Gamma^{R}_{\phantom a RR}=-\frac{2R}{(R^{2}+1)}\quad\quad\Gamma^{R}_{\phantom a \Phi\Phi}=\frac{R(R^{2}-1)}{(R^{2}+1)}\\ 
&&\quad\quad\quad\quad\Gamma^{\Phi}_{\phantom a R\Phi}=-\frac{(R^{2}-1)}{R(R^{2}+1)}.
\end{eqnarray*}
\noindent We now assume the following \textit{ansatz} for the map $\varphi(\eta,\theta,\psi)=(R(\eta,\theta,\psi), \Phi(\eta,\theta,\psi))$:
\begin{equation}\label{ansatz}
R=f(\eta)\quad\quad\quad \Phi=a\theta+b\psi 
\end{equation}
$a,b \in\mathbb{Z}$ being associated with angular windings around the two generating circles of the torus. Roughly, this implies that each torus in $\mathbb{R}^{3}$ is mapped into a parallel in $\mathbb{S}^{2}$. Adittionaly, in order to match the appropriate boundary conditions we require that $f(0)=0$ (meaning that spatial infinity i.e. the ``vacuum"\  is actually mapped into $\mathcal{N}$) and $f(\infty)=\infty$ (meaning that the soliton position is mapped into $\mathcal{S}$).

Now we note that the preimage of a given point in $\mathbb{S}^{2}$ is given by the equations
\begin{equation}
\eta=const_{1}\quad\quad\quad a\theta+b\psi=const_{2}\; , 
\end{equation}
which define non-intersecting closed loops winding around each torus. As is well known, see \cite{Manton}, the linking number between two of such arbitrary preimage curves is given by $Q=ab$ and uniquely determines the Hopf invariant for the solutions. Given $Q$, the collection of all preimage curves define a nontrivial fibration of $\mathbb{R}^{3}$ within the corresponding homotopy class. In terms of the above quantities, we get, for the first symmetric polynomial,
\begin{eqnarray}\label{Delta}
&&\sigma_{1}=\frac{4f^{2}q^{2}}{(1+f^{2})^{2}}\Delta^{2}\\
&&\Delta\equiv\big[(f^{'}/f)^{2}+(a^{2}+b^{2}\mbox{sinh}^{-2}\eta)\big]^{1/2}.
\end{eqnarray}
where $f^{'}\equiv\partial_{\eta}f$. With these assumptions, it is an easy task to see that Eq. (\ref{EOM}) is identically satisfied for $\Phi$ ({\em i.e.}, $\varphi^2$ in that equation) while the equation for $f$ ({\em i.e.}, $\varphi^1$) becomes the nonlinear ODE:
\begin{eqnarray}\label{profi}
&&\frac{(1+f^{2})}{\Delta\ \mbox{sinh}\eta\ f^{2}}\left[\frac{\Delta\ \mbox{sinh}\eta\ ff^{'}}{(1+f^{2})}\right]^{'}=2f^{'2}+\\\nonumber
&&\quad\quad+(1-f^{2})(a^{2}+b^{2}\mbox{sinh}^{-2}\eta)\; .
\end{eqnarray}
Finally, for the \textit{ansatz} provided by Eqs. (\ref{ansatz}) we get, using equation (\ref{reciprocal}), the reciprocal effective metric:
\begin{widetext}
\begin{equation}
(m^{-1})^{ij}=
q^{2}\left[\left( \begin{array}{ccc}
\ 1 & 0 & 0 \\
0 & 1 & 0 \\
0 & 0 & \mbox{sinh}^{-2}\eta \end{array} \right)+\frac{1}{\Delta^{2}}\left( \begin{array}{ccc}
\ (f^{'}/f)^{2} & 0 & 0 \\
0 & a^{2} & ab\ \mbox{sinh}^{-2}\eta \\
0 & ab\ \mbox{sinh}^{-2}\eta  & b^{2}\ \mbox{sinh}^{-4}\eta \end{array} \right)\right]
\end{equation}
\end{widetext}

\section{Geodesics about a $Q=1$ Hopfion}

In this section we present the results of numerical computations in the simpler case $Q=1$. In this case, it is possible to obtain the exact solution $f(\eta)=sinh(\eta)$ for the ``profile function" $f$ determined by (\ref{profi}), see \cite{Adam}, which uniquely determines the effective geometry. Equipped with this solution, we study geodesic motion in this geometric approach by considering the system (\ref{ODE}) and (\ref{CONS}). We integrate this  coupled system with a classical Runge-Kutta  method with automatic step-size control  for a large number of geodesics emerging from different regions in three-space. For computational reasons, we work here with cartesian coordinates $(x,y,z)$. The reciprocal effective metric then becomes (see the Appendix for the complete geodesic equations)\\
\begin{widetext}
 \begin{equation}\nonumber
(m^{-1})^{ij}= \left(
\begin{array}{ccc}
 \frac{3}{2}-\frac{2 (y-x z)^2}{\left(1+r^2\right)^2} & \frac{2 (y-x z) (x+y z)}{\left(1+r^2\right)^2} & -\frac{(y-x z) \left(x^2+y^2-z^2-1\right)}{\left(1+r^2\right)^2} \\
 \frac{2 (y-x z) (x+y z)}{\left(1+r^2\right)^2} & \frac{3}{2}-\frac{2 (x+y z)^2}{\left(1+r^2\right)^2} & \frac{(x+y z) \left(x^2+y^2-z^2-1\right)}{\left(1+r^2\right)^2} \\
 -\frac{(y-x z) \left(x^2+y^2-z^2-1\right)}{\left(1+r^2\right)^2} & \frac{(x+y z) \left(x^2+y^2-z^2-1\right)}{\left(1+r^2\right)^2} & \frac{x^4+2 x^2 \left(y^2+2 z^2+2\right)+y^4+4 y^2 \left(z^2+1\right)+\left(z^2+1\right)^2}{\left(1+r^2\right)^2} \\
\end{array}
\right)
\end{equation}
\end{widetext}
in which $r = \sqrt(x^2 + y^2 + z^2)$.

It can be checked that $(m^{-1})^{ij}$ is positive definite for all values of $(x,y,z)$ and is invariant under rotations about the $z-$axis. This last result is a direct consequence of the azymuthal symmetry of the underlying fibration. The fact that this metric is curved may be easily confirmed by calculating its Ricci scalar. Surprisingly, we obtain the simple result
\begin{equation}
\textbf{R}= -\frac{4 x^2+4 y^2-8 z^2+2}{\left(x^2+y^2+z^2+1\right)^2}\; , 
\end{equation} 
which is bounded and well-behaved everywhere. Interestingly, there are two non-connected surfaces in 3-space on which $\textbf{R}$ vanishes. They work as boundaries in three-dimensional space separating two qualitatively different regions, i.e. $\textbf{R}<0$ and $\textbf{R}>0$. Close to the core of the fibration ($x^2+y^2=1,\ z=0$), $\textbf{R}$ is negative and spheres have an excess of area compared to Euclidean spheres. Conversely, as one travels along the $z$-axis, $\textbf{R}$ becomes positive at some point and spheres appear to have a deficit of area compared to Euclidean spheres. A direct calculation shows that two global maxima exist for the points $(0,0,\pm\sqrt{3/2})$ while a global minima exist for the point $(0,0,0)$. In FIG. 1 it is depicted the level sets of the Ricci scalar, the core of the fibration (ring) and the preimage of the north pole (z-axis).
\begin{figure}[h]
       \centering  
       \includegraphics[scale=.35]{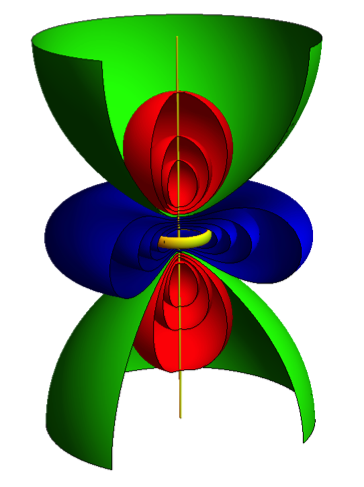}
       \caption{Level sets of the Ricci scalar for the $Q=1$ static Hopfion. Depicted in blue we have surfaces with $\textbf{R}<0$, while in red with $\textbf{R}>0$. Surfaces in Green represent the separatrix region with $\textbf{R}=0$. We have removed a wedge in order to reveal the internal structure of the sets.}
       \label{dynamic1}
\end{figure} 

Also, the static Hopfion scatter the rays nontrivially. The closer to the Hopfion, the greater the bending of rays - just like using denser materials to make optical lenses results in a greater amount of refraction. However, there is a striking novelty here: as the soliton carry a topological invariant, the effective metric also carries this number implicitly, `warping space' in a qualitatively different way for different values of $Q$. In this sense, different fibrations corresponding to different values of $Q$ will act as different nontrivial gravitational lenses for magnifying, distorting and refocusing distant `objects' (sources).

Let us investigate how effective geodesics behave in some specific situations. Starting with a `source' located at the point $(3,0,0)$ we have analysed several emanating geodesics with initial tangent vectors lying in the plane $z=0$ at time $t=0$. Solving the system (\ref{ODE}) and (\ref{CONS}) numerically for $n=314$ geodesics with $0<t<8$ we obtained the `disk' represented in FIG 2. 
\begin{figure}[h]
       \centering  
       \includegraphics[scale=.65]{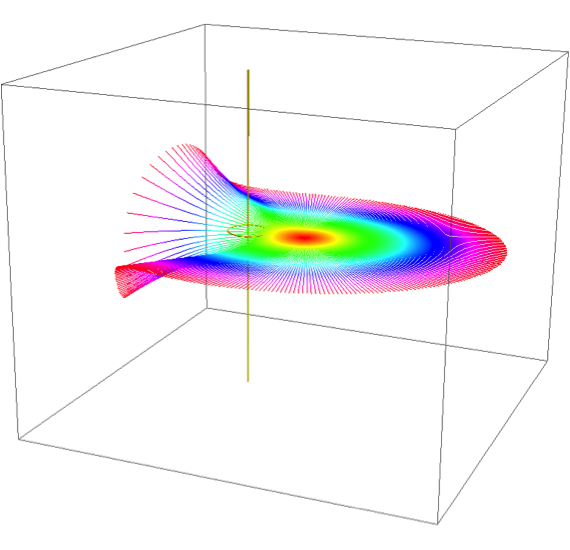}
       \caption{Geodesic disks centered at the point $(3,0,0)$. The captured process was made using the Runge-Kutta method, with time interval $0< t < 8$.  Colors represent disks of different radii associated to the values of the parameter $t$. }
       \label{dynamic1}
\end{figure} 
We see that, for sufficiently small times, the disk is nearly flat and tangent to the plane. The same is valid for geodesics travelling in the opposite direction of the Hopfion. However, when geodesics approach the core of the fibration, drastic distortions emerge, scattering some of them upwards and others downwards. This is in contrast with gravitational lensing in general relativity due to a spherically symmetric static mass. In the latter case, null geodesics initially lying in the plane stay in the plane forever.

Geometrically speaking, the disks presented in FIG. 2 all have an excess of area (and perimeter) as compared to Euclidean disks with the same radii. This result is expected since they live in the region $R<0$ which tends to defocus geodesic in the average, see FIG. 3.
\begin{figure}[h]
       \centering  
       \includegraphics[scale=.5]{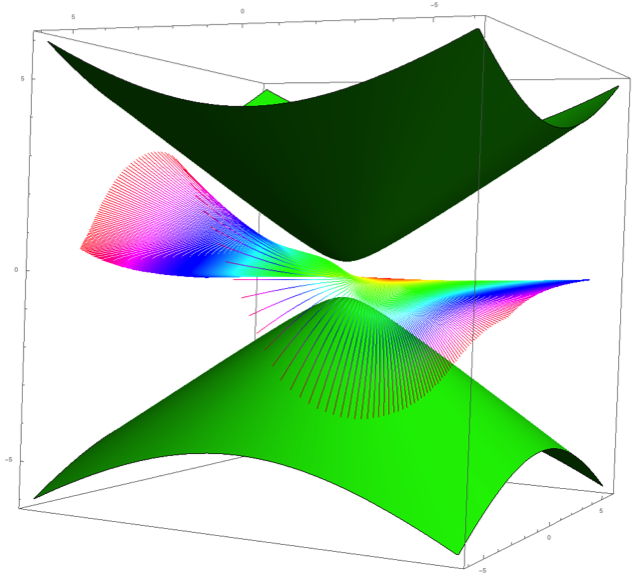}
       \caption{Defocusing of geodesics in the region with $R<1$ and separatrix (green) with $R=0$.}
       \label{dynamic1}
\end{figure} 

We have tested this defocusing effect for an extended unidimensional source characterised by $x=5, -5<y<5, z=0$. For this case the number of geodesics is $n=200$ while $0<t<15$. According to FIG. 4 similar results occur.
\begin{figure}[h]
       \centering  
       \includegraphics[scale=0.75]{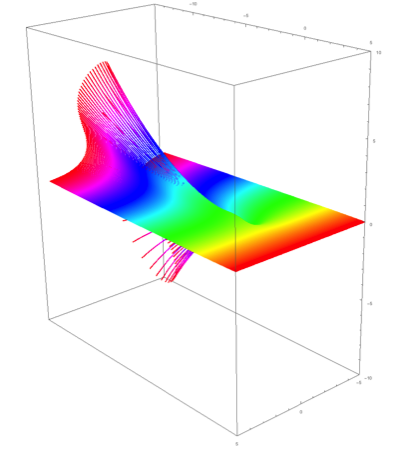}
       \caption{Scattering of incoming geodesics emanating from an extended unidimensional source of Euclidean length 10.  The captured process was made using the Runge-Kutta method, with time interval for $0<t<15$.}
       \label{dynamic2}
\end{figure} 
Another interesting example is the extended unidimensional source characterised by $x=5, y=0, -5<z<5$. This is effectively the same as to rotate by $\pi/2$ our last `experiment'. Here, the initial geodesics lie in the plane $y=0$ and are initially parallel. A subset of these geodesics are shown in FIG 5.

\begin{figure}[h]
       \centering  
       \includegraphics[scale=.6]{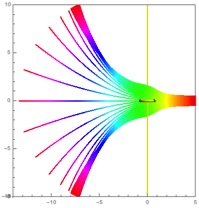}
       \caption{Scattering of incoming geodesics emanating from an extended unidimensional source. The captured process was made using the Runge-Kutta method for $0<t<15$.}
       \label{dynamic3}
\end{figure} 

Let us finally consider geodesics emanating from the region $\textbf{R}>0$. In this region, the curvature of three-dimensional space due to the Hopfion tends to focus the curves instead. Starting with geodesics emanating from the point $(0,0,5)$ and initially making an Euclidean angle of $\pi/4$ with the z-axis we see that they first converge, diverge for a while and converge again to meet at the point $(0,0,-5)$, see FIG. 6. Note that for these curves, the Hopfion behaves as a convergent lens for the effective geodesics.

\begin{figure}[h]
       \centering  
       \includegraphics[scale=.75]{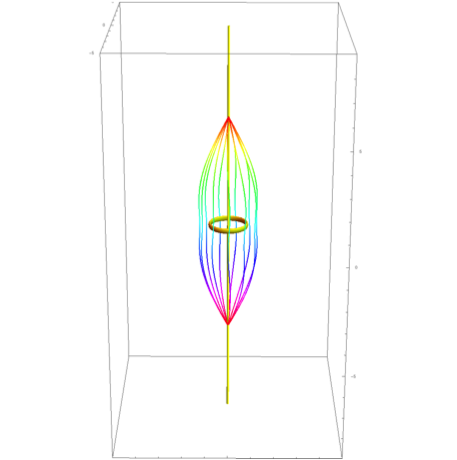}
       \caption{Pencil of 12 effective geodesics being focused by the Hopfion. Here we have considered geodesics emerging from the point (0,0,5) and having an initial angle of $\pi/4$ with the z-axis.}
       \label{dynamic3}
\end{figure} 

As a last example we consider a pencil of geodesics initially parallel to the z-axis. Interestingly, they pass through the core of the fibration yielding the pattern shown in FIG.7

\begin{figure}[h]
       \centering  
       \includegraphics[scale=.75]{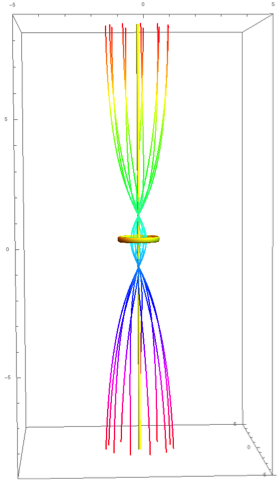}
       \caption{Pencil of 12 effective geodesics being focused by the Hopfion in an exotic way. Note that two distinct focal points emerge.}
       \label{dynamic3}
\end{figure}

\section{Conclusions}

The net result of this paper is the derivation of an effective metric description of wavy disturbances in the Nicole model. As is well known, this model is a non-polynomial generalization of the $O(3)$ $\sigma$-model and supports a Hopf-type fibration as an exact finite-energy solution to the nonlinear equations of motion. In Section III, we have discussed the hyperbolicity properties of the model and showed, in particular, that the evolution of high-frequency excitations about all possible static solutions is well-behaved. Roughly, this means that, as long as the background solutions are static, the linearized equations will entail a definite speed of propagation independently of spacetime position or direction of propagation. Section IV discusses the causal replacement of the model and shows that it is governed by a fouth-order Fresnel-like \textit{dispersion relation} given by (\ref{Fres}). Interestingly, this algebraic equation factorizes and gives rise to two pseudo-Riemannian effective metrics. One of these metrics is given in terms of the background map and depends implicitly on the topological invariant (Hopf index). This last result gives rise to an exotic structure for the causal replacement associated to the model. In order to explore our results further we have studied the behaviour of several effective geodesics about the exact Hopf fibration and showed that they behave in a quite unexpected way. As the mathematical structure of several theories supporting topological solitons is basically the same as that of the Nicole model, our results strongly suggest that similar effects are very likely to occur in effective field theories such as the Skyrme model of pions. It would be interesting to have any deeper insight into this last problem.

\section{Appendix}
Here we give explicit formulas for the geodesic equations and briefly summarize the Runge-Kutta method. In cartesian coordinates, $m_{ij}$ is given by the matrix
\begin{widetext}
 \begin{equation}\nonumber
\tiny{\left(
\begin{array}{ccc}
 \frac{2 \left(x^4+2 x^2 \left(y^2+2 z^2+1\right)-4 x y z+y^4+2 y^2
   \left(z^2+2\right)+\left(z^2+1\right)^2\right)}{3 \left(x^2+y^2+z^2+1\right)^2} &
   -\frac{4 (y-x z) (x+y z)}{3 \left(x^2+y^2+z^2+1\right)^2} & \frac{2 (y-x z)
   \left(x^2+y^2-z^2-1\right)}{3 \left(x^2+y^2+z^2+1\right)^2} \\
 -\frac{4 (y-x z) (x+y z)}{3 \left(x^2+y^2+z^2+1\right)^2} & \frac{2 \left(x^4+2 x^2
   \left(y^2+z^2+2\right)+4 x y z+y^4+y^2 \left(4
   z^2+2\right)+\left(z^2+1\right)^2\right)}{3 \left(x^2+y^2+z^2+1\right)^2} & -\frac{2
   (x+y z) \left(x^2+y^2-z^2-1\right)}{3 \left(x^2+y^2+z^2+1\right)^2} \\
 \frac{2 (y-x z) \left(x^2+y^2-z^2-1\right)}{3 \left(x^2+y^2+z^2+1\right)^2} & -\frac{2
   (x+y z) \left(x^2+y^2-z^2-1\right)}{3 \left(x^2+y^2+z^2+1\right)^2} & 1-\frac{4
   \left(z^2+1\right) \left(x^2+y^2\right)}{3 \left(x^2+y^2+z^2+1\right)^2} \\
\end{array}
\right)}
\end{equation}
\end{widetext}
which yields the geodesic equations
\begin{center}
\begin{widetext}
\tiny{
\begin{align}
&\frac{dx^2}{dt^2}-\frac{4}{3}\frac{1}{\left(1+{x}^{2}+{y}^{2}+{z}^{2}\right)^{4}}\left(\frac{}{}\right.-(x z+y) \left(z^3 \left(x^2+2 y^2+2\right)-3 x y \left(x^2+y^2+1\right)\right.\\\nonumber
&\qquad +z \left(-2 x^4-x^2 \left(y^2+1\right)+y^4+4 y^2+1\right)+x y z^2+\left.\left.z^5\right)\frac{}{}\right)\left(\frac{dx}{dt}\right)^{2}\\\nonumber
&\qquad +\frac{4}{3}\frac{1}{\left(1+{x}^{2}+{y}^{2}+{z}^{2}\right)^{4}}\left(\frac{}{}\right.-(x z+y) \left(-2 z^2 \left(x^2+2 y^2+3\right)+2 x y z \left(3 x^2+3 y^2+5\right)\right.\\\nonumber
&\qquad \left.-3 \left(x^2+1\right)^2+2 x y z^3+3 y^4-3 z^4\right)\left(\frac{dx}{dt}\right)\left(\frac{dy}{dt}\right)\\\nonumber
&\qquad +\frac{4}{3}\frac{1}{\left(1+{x}^{2}+{y}^{2}+{z}^{2}\right)^{4}}\left(\frac{}{}\right.(x z+y) \left(-2 x z^2 \left(x^2+y^2-1\right)-4 y z \left(2 x^2+2 y^2+1\right)\right.\\\nonumber
&\qquad  \left.+3 x \left(x^2+y^2+1\right)^2-x z^4-4 y z^3\right)\left(\frac{dx}{dt}\right)\left(\frac{dz}{dt}\right)\\\nonumber
&\qquad +\frac{4}{3}\frac{1}{\left(1+{x}^{2}+{y}^{2}+{z}^{2}\right)^{4}}\left(\frac{}{}\right. x z^4 \left(2 x^2+y^2-1\right)+4 y z^3 \left(x^2+y^2+2\right)\\\nonumber
&\qquad +3 x \left(y^2-1\right) \left(x^2+y^2+1\right)+x z^2 \left(x^4-\left(x^2+5\right) y^2+x^2-2 y^4-5\right)\\\nonumber
&\qquad +2 y z\left(2 x^4+x^2 \left(y^2+5\right)-y^4+y^2+2\right)+x z^6+4 y z^5\left.\frac{}{}\right)\left(\frac{dy}{dt}\right)^{2}\\\nonumber
&\qquad -\frac{2}{3}\frac{1}{\left(1+{x}^{2}+{y}^{2}+{z}^{2}\right)^{4}}\left(\frac{}{}\right. 3 x^6-6 x^5 y z+x^4 \left(3 y^2-z^2+15\right)+4 x^3 y z \left(-3 y^2+z^2-7\right)\\\nonumber
&\qquad  +x^2 \left(-3 y^4+18 y^2 \left(z^2+1\right)+z^4+10 z^2+9\right)-2 x y z \left(3 y^4-2 y^2\left(z^2-7\right)-z^4+6 z^2+7\right)\\\nonumber
&-\qquad 3 y^6+y^4 \left(19 z^2+3\right)+y^2 \left(z^2+1\right) \left(11 z^2+3\right)-3 \left(z^2+1\right)^3\left.\frac{}{}\right)\left(\frac{dy}{dt}\right)\left(\frac{dz}{dt}\right)\\\nonumber
&\qquad -\frac{2}{3}\frac{1}{\left(1+{x}^{2}+{y}^{2}+{z}^{2}\right)^{4}}\left(\frac{}{}\right.-x z^4 \left(7 x^2+7 y^2+15\right)-4 y z^3 \left(x^2+y^2-3\right)\\\nonumber
&\qquad +3 x \left(x^2+y^2-1\right) \left(x^2+y^2+1\right) \left(x^2+y^2+3\right)-x z^2 \left(9 x^4+2 x^2 \left(9y^2+5\right)+9 y^4+10 y^2+21\right)\\\nonumber
&\qquad -2 y z \left(9 x^4+2 x^2 \left(9 y^2+1\right)+9 y^4+2 y^2-3\right)-3 x z^6+6 y z^5\left.\frac{}{}\right)\left(\frac{dz}{dt}\right)^{2}=0 \; , \\\\\nonumber
&\frac{dy^2}{dt^2}+\frac{4}{3}\frac{1}{\left(1+{x}^{2}+{y}^{2}+{z}^{2}\right)^{4}}\left(\frac{}{}\right. y z^4 \left(x^2+2 y^2-1\right)-4 x z^3 \left(x^2+y^2+2\right)\\\nonumber
&\qquad +3 \left(x^2-1\right) y \left(x^2+y^2+1\right)+y z^2 \left(-2 x^4-x^2 \left(y^2+5\right)+y^4+y^2-5\right)\\\nonumber
&\qquad +2 x z\left(x^4-\left(x^2+5\right) y^2-x^2-2 y^4-2\right)-4 x z^5+y z^6\left.\frac{}{}\right)\left(\frac{dx}{dt}\right)^{2}\\\nonumber
&\qquad -\frac{4}{3}\frac{1}{\left(1+{x}^{2}+{y}^{2}+{z}^{2}\right)^{4}}\left(\frac{}{}\right.(x-y z) \left(3 x^4-2 z^2 \left(2 x^2+y^2+3\right)\right.\\\nonumber
&\qquad \left.-2 x y z \left(3 x^2+3 y^2+5\right)-2 x y z^3-3 \left(y^2+1\right)^2-3 z^4\right)\left(\frac{dx}{dt}\right)\left(\frac{dy}{dt}\right)\\\nonumber
&\qquad -\frac{2}{3}\frac{1}{\left(1+{x}^{2}+{y}^{2}+{z}^{2}\right)^{4}}\left(\frac{}{}\right.3 x^6-6 x^5 y z+x^4 \left(3 y^2-19 z^2-3\right)+4 x^3 y z \left(-3 y^2+z^2-7\right)\\\nonumber
&\qquad -x^2 \left(3 y^4+18 y^2 \left(z^2+1\right)+11 z^4+14 z^2+3\right)-2 x y z \left(3 y^4-2 y^2\left(z^2-7\right)-z^4+6 z^2+7\right)\\\nonumber
&\qquad -3 y^6+y^4 \left(z^2-15\right)-y^2 \left(z^2+1\right) \left(z^2+9\right)+3 \left(z^2+1\right)^3\left.\frac{}{}\right)\left(\frac{dx}{dt}\right)\left(\frac{dz}{dt}\right)\\\nonumber
&\qquad -\frac{4}{3}\frac{1}{\left(1+{x}^{2}+{y}^{2}+{z}^{2}\right)^{4}}\left(\frac{}{}\right. (x-y z) \left(z^3 \left(2 x^2+y^2+2\right)+3 x y \left(x^2+y^2+1\right)\right.\\\nonumber
&\qquad +\left.z \left(x^4-\left(x^2+1\right) y^2+4 x^2-2 y^4+1\right)-x y z^2+z^5\right)\left(\frac{dy}{dt}\right)^{2}\\\nonumber
&\qquad -\frac{4}{3}\frac{1}{\left(1+{x}^{2}+{y}^{2}+{z}^{2}\right)^{4}}\left(\frac{}{}\right.(x-y z) \left(-2 y z^2 \left(x^2+y^2-1\right)+4 x z \left(2 x^2+2 y^2+1\right)\right.\\\nonumber
&\qquad +\left.3 y \left(x^2+y^2+1\right)^2+4 x z^3-y z^4\right)\left(\frac{dy}{dt}\right)\left(\frac{dz}{dt}\right)\\\nonumber
&\qquad -\frac{1}{3}\frac{1}{\left(1+{x}^{2}+{y}^{2}+{z}^{2}\right)^{4}}\left(\frac{}{}\right.-y z^4 \left(7 x^2+7 y^2+15\right)+4 x z^3 \left(x^2+y^2-3\right)\\\nonumber
&\qquad +3 y \left(x^2+y^2-1\right) \left(x^2+y^2+1\right) \left(x^2+y^2+3\right)-y z^2 \left(9 x^4+2 x^2 \left(9y^2+5\right)+9 y^4+10 y^2+21\right)\\\nonumber
&\qquad +2 x z \left(9 x^4+2 x^2 \left(9 y^2+1\right)+9 y^4+2 y^2-3\right)-6 x z^5-3 y z^6\left.\frac{}{}\right)\left(\frac{dz}{dt}\right)^{2}=0 \; ,
\end{align}}
\end{widetext}
\end{center}

\begin{center}
\begin{widetext}
\tiny{
\begin{align}
&\frac{dz^2}{dt^2}-\frac{2}{3}\frac{1}{\left(1+{x}^{2}+{y}^{2}+{z}^{2}\right)^{4}}\left(\frac{}{}\right. x^6 z+x^4 z \left(3 y^2+9 z^2+13\right)+4 x^3 y \left(z^2+3\right)\\\nonumber
&\qquad +x^2 z \left(3 y^4+10 y^2 \left(z^2+1\right)+3 z^4+14 z^2+11\right)+4 x y \left(y^2 \left(z^2+3\right)-z^4+2z^2+3\right)\\\nonumber
&\qquad +z \left(y^6+y^4 \left(z^2-3\right)-y^2 \left(z^2+1\right) \left(z^2+9\right)-\left(z^2+1\right)^3\right)\left(\frac{dx}{dt}\right)^{2}\\\nonumber
&\qquad +\frac{8}{3}\frac{1}{\left(1+{x}^{2}+{y}^{2}+{z}^{2}\right)^{4}}\left(\frac{}{}\right. x^4 \left(z^2+3\right)-4 x^3 y z \left(z^2+2\right)+x^2 \left(-z^4+2 z^2+3\right)\\\nonumber
&\qquad -2 x y z \left(2 y^2 \left(z^2+2\right)+z^4+6 z^2+5\right)-y^2 \left(y^2\left(z^2+3\right)-z^4+2 z^2+3\right)\left.\frac{}{}\right)\left(\frac{dx}{dt}\right)\left(\frac{dy}{dt}\right)\\\nonumber
&\qquad +\frac{4}{3}\frac{1}{\left(1+{x}^{2}+{y}^{2}+{z}^{2}\right)^{4}}\left(\frac{}{}\right. \left(x^2+y^2-z^2-1\right) \left(x z^2 \left(4 x^2+4 y^2+5\right)\right.\\\nonumber
&\qquad +\left.y z \left(x^2+y^2-1\right)+3 x \left(x^2+y^2+1\right)+2 x z^4-y z^3\right)\left(\frac{dx}{dt}\right)\left(\frac{dz}{dt}\right)\\\nonumber
&\qquad -\frac{2}{3}\frac{1}{\left(1+{x}^{2}+{y}^{2}+{z}^{2}\right)^{4}}\left(\frac{}{}\right. x^6 z+x^4 z \left(3 y^2+z^2-3\right)-4 x^3 y \left(z^2+3\right)\\\nonumber
&\qquad -x^2 z \left(-3 y^4-10 y^2 \left(z^2+1\right)+z^4+10 z^2+9\right)-4 x y \left(y^2 \left(z^2+3\right)-z^4+2z^2+3\right)\\\nonumber
&\qquad +z \left(y^6+y^4 \left(9 z^2+13\right)+y^2 \left(z^2+1\right) \left(3 z^2+11\right)-\left(z^2+1\right)^3\right)\left.\frac{}{}\right)\left(\frac{dy}{dt}\right)^{2}\\\nonumber
&\qquad -\frac{4}{3}\frac{1}{\left(1+{x}^{2}+{y}^{2}+{z}^{2}\right)^{4}}\left(\frac{}{}\right.(-1 + x^2 + y^2 - z^2) (-3 y (1 + x^2 + y^2) + x (-1 + x^2 + y^2) z\\\nonumber
&\qquad -y (5 + 4 x^2 + 4 y^2) z^2 - x z^3 - 2 y z^4)\left.\frac{}{}\right)\left(\frac{dy}{dt}\right)\left(\frac{dz}{dt}\right)\\\nonumber
&\qquad -\frac{2}{3}\frac{1}{\left(1+{x}^{2}+{y}^{2}+{z}^{2}\right)^{4}}\left(z \left(x^2+y^2\right) \left(x^2+y^2-z^2-1\right) \left(3 x^2+3 y^2+z^2+1\right)\right)\left(\frac{dz}{dt}\right)^{2}=0 \; .
\end{align}}
\end{widetext}
\end{center}
In order to solve these equations numerically we recall that a system of ODE of order  $n,$
\begin{equation}
\frac{d^{n}y}{dt^{n}}=f\left(t,y,y',\ldots,\frac{dy^{n-1}}{dt}\right),
\label{ODE_N}
\end{equation}
can be reduced to a first order system as follows. Let $u=(u_{0},\ldots,u_{n-1}),$ with
\begin{align}
u_{0}=y,\,\,\,u_{j}=\frac{dy^{j}}{dt},
\label{sustitucion}
\end{align}
where $j=1,\ldots,n-1.$ Then, $\eqref{ODE_N}$ takes the form
\begin{equation}\nonumber
\frac{du}{dt}=(u_{1},\ldots,u_{n-1},f(t,u_{0},\ldots,u_{n-1}))=g(t,u).
\end{equation}
If $y$ take values in $\mathbb{R}^{k},$ then $u$ take values in $\mathbb{R}^{kn}.$ In our case $n=2$ and we have
 \begin{equation}
 \left.
\begin{array}{ll}
u'(t)  = g(t,u(t)),  \\
u(t_{0})=u_0 
\end{array}
\right\rbrace, t\in[t_{0},t_{f}]. 
\end{equation}
We then consider a partition of the interval $[t_{0},t_{f}],$ i.e. $$t_{0}<t_{1}<\ldots<t_{N}=t_{f}$$ and the corresponding approximations $u_{n}\approx u(t_{n}),$ with $n=1,2,\ldots,N.$ Then, the explicit \textit{Runge}-\textit{Kutta} methods of $s$  stages are 
\begin{align}
\begin{split}
  k_{1} &= g(t_{n},u_{n}) \\
  k_{2} &= g(t_{n}+c_{2}h_{n},u_{n} + h_{n}a_{21}k_{1}) \\
  k_{3} &= g(t_{n}+c_{3}h_{n},u_{n} + h_{n}(a_{31}k_{1}+a_{32}k_{2})) \\
   \vdots &=\qquad\qquad\vdots\\
  k_{s} &= g(t_{n}+c_{s}h_{n},u_{n} + h_{n}(a_{s1}k_{1} + \ldots + a_{s,s-1}k_{s-1}))\\
  u_{n+1} &= u_{n} + \ h_{n}(b_{1}k_{1}+\ldots+b_{s}k_{s}).
\end{split}
\end{align}
This method can be written as \textit{Butcher tableau} 
\begin{center}
   \begin{tabular}{ c | c c  c c c}
     0          & 0          &              &            &                 & \\ 
     $c_{2}$   & $a_{21}$ &$ a_{32}$ &            &                 &\\ 
     $\vdots$ & $\vdots $ & $\vdots $ &$\ddots$  &                  & \\
     $c_{s}$   & $a_{s1} $ & $a_{s2} $ & $\ldots$ & $a_{s,s-1}$ & \\\hline
                      & $b_{1}$    & $b_{2}$  & $\ldots$   & $b_{s-1}$    & $bs$ 
   \end{tabular}
 \end{center}
In our case,  we have used the classic  Runge-Kutta method, i.e.
\begin{center}
   \begin{tabular}{ c | c c  c c c}
     0          & 0          &              &            &                 & \\ 
     $\frac{1}{2}$   & $\frac{1}{2}$ &  &            &                 &\\ 
     $\frac{1}{2}$ & $0 $ & $\frac{1}{2} $ &  &                  & \\
     $1$   & $0 $ & $0 $ & $1$ &  & \\\hline
                      & $\frac{1}{6}$    & $\frac{1}{3}$  & $\frac{1}{3}$   & $\frac{1}{6}$    & 
   \end{tabular}.
 \end{center}

 \textbf{Acknowledgements} E.G. would like to thank FAPESP (grant 2011/11973-4) for funding his visit to ICTP-SAIFR from 08-09 2015 where part of this work was done, and CAPES for financial support.

M.B. would like to thank Becas Chile (Concurso Becas de Doctorado en el Extranjero) for financial support. 

E.G.R.'s work was partially supported by the project FONDECYT \# 1161691.

\end{document}